\documentclass[a4paper,11pt]{article}
\pdfoutput=1 
\usepackage{jheppub} 
\usepackage[T1]{fontenc} 
\usepackage{color}

\def \be {\begin{equation}}
\def \ee {\end{equation}}
\def \ba {\begin{aligned}}
\def \ea {\end{aligned}}
\def \bea {\begin{eqnarray}}
\def \eea {\end{eqnarray}}

\title{Stochastic Thermodynamics of Black Hole Phase Transition in Extended Phase Space}

\author[a]{Conghua Liu,}
\author[b,*]{and Jin Wang \note[*]{Corresponding author.}}
\affiliation[a]{College of Physics, Jilin University, Changchun 130022, China}
\emailAdd{liuch2026@jlu.edu.cn}
\affiliation[b]{Department of Chemistry and Department of Physics and Astronomy, State University of New York at Stony Brook, Stony Brook, New York 11794, USA}
\emailAdd{jin.wang.1@stonybrook.edu}

\abstract{Black holes provide one of the most profound arenas in which gravity and thermodynamics meet. In the extended phase space, they further exhibit phase transitions, criticality, and heat-engine cycles that closely resemble those of ordinary thermodynamic matter. Yet these behaviors have so far been formulated mainly at the macroscopic or ensemble level, leaving their fluctuations and irreversibility largely unexplored. Here, we present the first stochastic thermodynamic formulation of the black hole phase transitions in the extended phase space, grounded in a fully thermodynamic perspective and centered on their dynamical processes where fluctuations play a dominant role. We derive an integral fluctuation theorem and a Jarzynski-like equality, where the total entropy production can be identified as a sum of three contributions: the change in information entropy, the change in Bekenstein-Hawking entropy, and the entropy variation of the thermal bath.  Moreover, we construct a stochastic Stirling black hole heat engine, whose efficiency can exceed the Carnot bound at the single-trajectory level. These theoretical predictions are validated through numerical simulations. Our framework establishes a direct bridge between conventional black hole thermodynamics and stochastic thermodynamics, offering new insights into black hole physics in the extended phase space.}

\begin{document}

\maketitle

\flushbottom

\section{Introduction}
Since the seminal insights of Bekenstein and Hawking, black holes have been recognized as thermodynamic systems possessing temperature and entropy, and obeying the laws of thermodynamics~\cite{AA,AB,AC,AD}. Over the past fifteen years, this framework has been substantially advanced through the development of extended phase space thermodynamics~\cite{AE,AF,AG,AH}, also known as black hole chemistry~\cite{AI, AJ}, in which the negative cosmological constant of Anti-de Sitter (AdS) spacetime is treated as a dynamical variable related to thermodynamic pressure. This reinterpretation not only yields well-defined equations of state, but also reveals a rich spectrum of phase structures and critical phenomena across diverse black hole solutions~\cite{AK,AL,AM,AN}. 

Concurrently, over the past three decades, stochastic thermodynamics has emerged as a foundational framework for systems with significant fluctuations~\cite{AO,AP}. It formulates thermodynamic quantities such as entropy production at the level of individual trajectories and establishes universal results, including fluctuation theorems~\cite{AQ}, the Jarzynski equality~\cite{AR}, and the fluctuation-dissipation theorem~\cite{AS}. Although stochastic thermodynamics was developed largely for mesoscopic systems, its basic ingredients---fluctuations and irreversibility---are not foreign to gravitational physics. Hawking radiation reflects the thermal quantum nature of horizons, and smaller black holes, including primordial black holes in the early universe, can display much stronger thermal and quantum fluctuations than astrophysical black holes~\cite{AA,AB,AC,AD}. At even larger cosmological scales, studies of eternal inflation and the multiverse have employed stochastic master or diffusion equations to describe vacuum transitions, bubble nucleation, and the evolution of probability distributions across different vacua~\cite{GGB,GGC,GGD,GGE}. The parallel developments of black hole thermodynamics and stochastic thermodynamics motivate a fundamental and compelling question: what new insights might their synthesis reveal? 

In recent years, the dynamics of black hole phase transitions have been described as a stochastic process governed by the Fokker-Planck equation~\cite{AT,AU}. Within this framework, thermal fluctuations play a pivotal role by enabling a stable black hole state to overcome the potential barrier provided by the generalized free energy and transit to another stable state, as shown in Fig.~\ref{Fig1}(a). This perspective has been further extended to a broad class of black hole systems~\cite{AV, AW, AX, AY, AZ, BA, BB, BC, BBBBA,ZZZZN,ZZZZO,ZZZZP,ZZZZQ} and admits an equivalent description in terms of Langevin dynamics~\cite{BD, BE, BF, BG}. Here, the generalized free energy landscape extends the usual on-shell thermodynamic description to off-shell black-hole configurations, where the Hawking temperature need not match the ensemble temperature and conical singularities generically arise. It has a solid basis from Hawking's gravitational path integral~\cite{BH,BG} and the Iyer-Wald formalism~\cite{BI}, and has also been used to reveal the thermodynamic topological properties of black holes~\cite{BJ,BK,BL,BM,BN,BO,BP,BQ,BR}.

The presence of significant fluctuations for black hole phase transitions at the dynamical level provides a natural starting point for formulating a stochastic thermodynamic description within black hole thermodynamics. Here, we establish a complete stochastic thermodynamic framework and show that the conventional second law, formulated in terms of geometric and environmental entropy, is not guaranteed to remain nonnegative for black hole phase transitions. A consistent thermodynamic description requires the inclusion of an information-theoretic entropy, leading to a generalized second law, whose nonnegativity is ensured by the integral fluctuation theorem. For processes connecting initial and final equilibrium states, we establish a Jarzynski-like equality. Furthermore, we extend conventional AdS black-hole heat engines into the stochastic regime by constructing a stochastic Stirling cycle, in which the efficiency can exceed the Carnot bound at the single-trajectory realization.


\section{Generalized free energy landscape and Langevin equation}

We consider a four-dimensional charged AdS black hole with metric
\be
\ba
\label{eq1}
ds^2=-f(R)dt^2+\frac{dR^2}{f(R)}+R^2d\Omega^2,
\ea
\ee
where $f(R)=1-{2M}/{R}+{q^2}/{R^2}+{R^2}/{l^2}$. Here, $M$, $q$, and $l$ represent the black hole mass, charge, and AdS curvature radius, respectively. In the extended phase space, the cosmological constant $\Lambda$ is identified as the thermodynamic pressure $P=-{\Lambda}/{8\pi}$~\cite{AE,AF,AG,AH}, with ${\Lambda}=-{3}/{l^2}$.

\begin{figure}[t]
\includegraphics[width=0.8\textwidth]{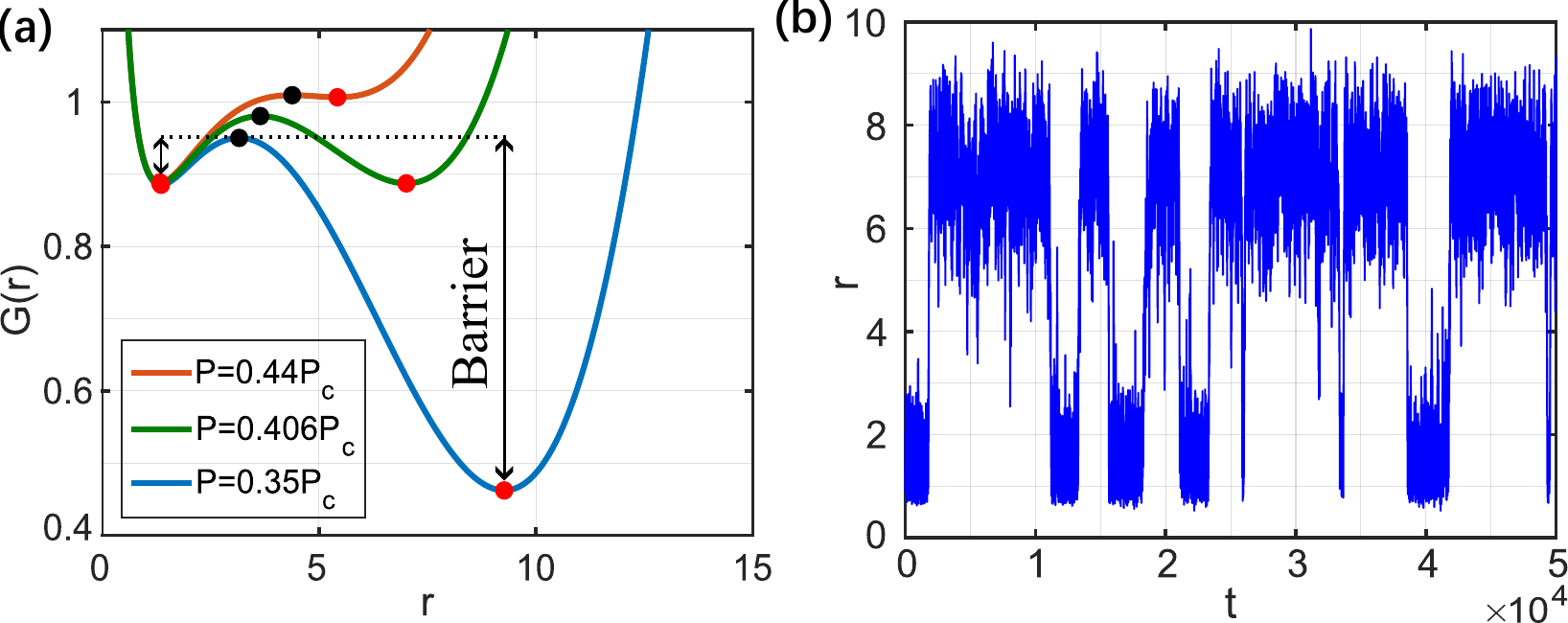}
\centering
\caption{(a) Generalized free energy landscape at different pressures for $q=1$ and $T_{\mathrm{E}}=0.03$, where the critical pressure is $P_c=1/96\pi q^2$. The double-headed arrows indicate the free energy barriers. The red (black) spheres denote locally stable (unstable) on-shell states, precisely corresponding to extrema of the free energy with $T_{\mathrm{E}}=T_{\mathrm{H}}$, while all other points on the curve represent off-shell states with $T_{\mathrm{E}}\neq T_{\mathrm{H}}$. (b) A representative stochastic trajectory of the horizon radius $r$ at $q=1$, $T_{\mathrm{E}}=0.03$, and $P=0.4P_c$.}
\label{Fig1}
\end{figure}

The generalized free energy of the charged AdS black hole is given by~\cite{AU,BF,BH}: $G=U+PV-T_{\text{E}}S_{\text{BH}}=r/2+q^2/2r+4\pi Pr^3/3-T_{\text{E}}\pi r^2$, where $U$ is the internal energy, $V$ is the thermodynamic volume, $S_{\rm BH}$ is the Bekenstein-Hawking entropy, and $r$ denotes the horizon radius. $T_{\text{E}}$ is the ensemble temperature, a free parameter that does not necessarily equal the Hawking temperature $T_{\text{H}}$. This expression can be rigorously derived from the Euclidean gravitational path integral by incorporating the off-shell contributions in the Einstein-Hilbert action~\cite{BH}; details are not repeated here. The corresponding generalized free energy landscape is shown in Fig.~\ref{Fig1} (a). Transitions between the two locally stable on-shell states (red spheres) require overcoming the free-energy barrier separating the initial stable state and the unstable on-shell state (black sphere), rendering thermal fluctuations essential.

Such stochastic dynamics of black hole phase transitions can be described by the Langevin equation~\cite{BD, BE, BF, BG}
\be
\ba
\label{eq2}
\frac{dr}{dt}=-D\beta_E\,\partial_r G(r,\lambda)+\eta(t),
\ea
\ee
where the generalized free energy provides the drift force and the thermal fluctuations provide the stochastic force. Here, $\beta_{\text{E}}=1/T_{\text{E}}$, $D$ is the diffusion coefficient, $\eta(t)$ is Gaussian white noise with correlation  $\langle\eta(t)\eta(t')\rangle=2D\,\delta(t-t')$, and $\lambda$ denotes an external control parameter. In the extended phase space, we choose the thermodynamic pressure \(P\), associated with the cosmological constant \(\Lambda\), as the control parameter. A representative stochastic trajectory of $r$ is shown in Fig.~\ref{Fig1}(b), illustrating that the horizon radius---and hence related thermodynamic quantities such as the internal energy and entropy---exhibit intrinsic stochasticity. Hereafter, we set $Q=1$ and $D=0.03$.

\section{Integral fluctuation theorem and second law}

The Langevin dynamics can be equivalently formulated within the Onsager-Machlup path-integral framework~\cite{BS,BT}, in which the conditional probability of a stochastic trajectory is expressed as: 
\be
\ba
p_{\text{F}}(r_{\text{F}}(t)|r_i)=N \exp\{-\int_{0}^{t_f}[\frac{1}{4D}{(\frac{dr}{dt}+\frac{D\beta_{\text{E}} \partial G(r,\lambda)}{\partial r})^2}-\frac{D\beta_{\text{E}} \partial^2 G(r,\lambda)}{2\partial r^2}]dt\}.
\ea
\ee
Here, $N$ denotes a normalization factor, and $p_{\text{F}}(r_{\text{F}}(t)|r_i)$ represents the conditional probability of a forward trajectory $r_{\text{F}}(t)$ starting from a fixed initial state $r_i$ in the forward process.

To define the backward process, its initial distribution is chosen to coincide with the final distribution of the forward process. The control parameter is time-reversed as $\tilde{\lambda}(t)=\lambda(t_f-t)$, and the backward trajectory $r_{\text{B}}(t)$ is constructed by time-reversing the state in the forward trajectory as $\tilde{r}(t)=r(t_f-t)$. Substituting the transformed variables $\{{r(t),\lambda(t)}\}\to\{{\tilde{r}(t),\tilde{\lambda}(t)}\}$ into the expression of the forward conditional probability, changing the integration variable $u=t_f-t$, and relabeling $u\to t$, we obtain the conditional probability of the backward trajectory in the backward process as \be
\ba
p_{\text{B}}(r_{\text{B}}(t)|r_f)=N \exp\{-\int_{0}^{t_f}[\frac{1}{4D}{(-\frac{dr}{dt}+\frac{D\beta_{\text{E}} \partial G(r,\lambda)}{\partial r})^2}-\frac{D\beta_{\text{E}} \partial^2 G(r,\lambda)}{2\partial r^2}]dt\}.
\ea
\ee

Taking the logarithmic ratio of the forward and backward conditional trajectory probabilities, we obtain
\be
\ba
\label{eq3}
\ln\frac{p_{\text{F}}(r_{\text{F}}(t)|r_i)}{p_{\text{B}}(r_{\text{B}}(t)|r_f)}&=-\frac{1}{T_{\text{E}}}\int_{0}^{t_f}\frac{\partial G(r,\lambda)}{\partial r}\frac{dr}{dt} dt\\
&=-\frac{1}{T_{\text{E}}}\int_{r_i}^{r_f}(\frac{1}{2}-\frac{q^2}{2r^2}+4\pi Pr^2-2T_{\text{E}}\pi r) dr.
\ea
\ee
Using the standard expression for the Hawking temperature $T_{\text{H}}=1/4\pi r-q^2/4\pi r^3+2Pr$, Eq.~(\ref{eq3}) can be recast in the thermodynamic form 
\be
\ba
\label{eq4}
\ln\frac{p_{\text{F}}(r_{\text{F}}(t)|r_i)}{p_{\text{B}}(r_{\text{B}}(t)|r_f)}
=-\frac{1}{T_{\text{E}}}\int T_{\text{H}} dS_{\text{BH}}+\Delta S_{\text{BH}}.
\ea
\ee

The distinction between the expressions for the on-shell and off-shell free energies can be viewed as the replacement of the Hawking temperature by the ensemble temperature~\cite{AT,AU}. From the thermodynamic identity $U=\partial_{\beta}\!\big[\beta(G-PV)\big]$, the internal energy of the off-shell black hole remains $U=r/2+q^2/2r$. The heat absorbed by the black hole is then determined from the energy conservation law as $\delta Q=dU-\delta W=(1/2-q^2/2r^2+4\pi Pr^2)dr=T_{\text{H}}dS_{\text{BH}}$, where $\delta W=-PdV$ denotes the work performed on the black hole. The change in entropy of the heat bath at the ensemble temperature $T_{\text{E}}$ is then given by $\Delta S_{\text{Bath}}=-\frac{1}{T_{\text{E}}}\int \delta Q=-\frac{1}{T_{\text{E}}}\int T_{\text{H}} dS_{\text{BH}}$. Collecting these results gives the ratio of the forward- and backward-path conditional probabilities:
\be
\ba
\label{eq5}
\ln\frac{p_{\text{F}}(r_{\text{F}}(t)|r_i)}{p_{\text{B}}(r_{\text{B}}(t)|r_f)}=\Delta S_{\text{Bath}}+\Delta S_{\text{BH}}.
\ea
\ee

The total entropy production $\sigma$ quantifies the time irreversibility of the trajectory, which is expressed as the logarithm of the ratio of forward and backward trajectory probabilities:
\be
\ba
\label{eq6}
\sigma=\ln\frac{p_{\text{F}}(r_{\text{F}}(t))}{p_{\text{B}}(r_{\text{B}}(t))}=\Delta S_{\text{Info}}+\Delta S_{\text{BH}}+\Delta S_{\text{Bath}},
\ea
\ee
where the change of information entropy is given by $\Delta S_{\text{Info}}=\ln p(r_i)-\ln p(r_f)$. The total entropy production satisfies the integral fluctuation theorem:
\be
\ba
\label{eq7}
\langle \exp(-\sigma)\rangle=\sum_{r_{\text{F}}(t)}p_{\text{F}}(r_{\text{F}}(t))\frac{p_{\text{B}}(r_{\text{B}}(t))}{p_{\text{F}}(r_{\text{F}}(t))}=1.
\ea
\ee
Using the Jensen inequality $\langle \exp(-\sigma)\rangle\geq \exp(-\langle\sigma\rangle)$, we can obtain the nonnegativity of the average entropy production:
\be
\ba
\label{eq8}
\langle \sigma\rangle=\langle \Delta S_{\text{Info}}+\Delta S_{\text{BH}}+\Delta S_{\text{Bath}}\rangle \geq 0.
\ea
\ee

Our analysis reveals that the conventional thermodynamic entropy change associated with the black hole and its environment, $\langle \Delta S_{\text{BH}}+\Delta S_{\text{Bath}}\rangle$, is not guaranteed to remain nonnegative during the black hole phase transition. In contrast, the full entropy production, $\langle \sigma\rangle$, which includes the information-theoretic contribution, is strictly nonnegative due to the integral fluctuation theorem. This reflects the distinct roles of the two entropy contributions: the gravitational entropy characterizes the intrinsic entropy of a given black hole configuration, whereas the information entropy quantifies the statistical weight over different configurations. Both contributions are therefore essential for capturing the thermodynamic irreversibility and formulating the second law. It is also worth noting that, except for the reversible limit where
\(\sigma=0\) for all trajectories, the integral fluctuation theorem requires the existence of trajectories with negative stochastic entropy
production \((\sigma<0)\); otherwise, one would have
\(\langle\exp(-\sigma)\rangle<1\).

\section{The Jarzynski-like equality}

Using the energy conservation law $\Delta U=W+Q$ and the relation $Q=-T_{\text{E}}\Delta S_{\text{Bath}}$, with $W$ and $Q$ being the work performed on and the heat absorbed by the black hole, respectively, Eq.~(\ref{eq7}) can 
be rewritten as $\langle\exp(-\sigma)\rangle
=\langle \exp[(\Delta U- W)/T_{\text{E}}-\Delta S_{\text{BH}}-\Delta S_{\text{Info}}]\rangle
=1$. We further assume that the initial and final states are in equilibrium, whose probability distribution is given by $p_{\text{eq}}=\exp(-\beta_{\text{E}} G)/Z$, where $Z$ is given by $Z=\int_0^{\infty}\exp(-\beta_{\text{E}} G)dr$~\cite{SSIEQ}. The initial and final distributions are not necessarily the same due to the difference in the external control parameter $\lambda$. The equilibrium distribution results from solving the corresponding Fokker-Planck equation $\frac{\partial p}{\partial t}=-\frac{\partial}{\partial r} J(r)=\frac{\partial}{\partial r}(\frac{G'(r)}{\gamma}p+D\frac{\partial p}{\partial r})=0$, where the steady state is an equilibrium state means that the steady-state probability flux $J(r)=-G'(r)p/\gamma-D\partial p/\partial r=0$.

The information entropy term can then be expressed as $-T_{\text{E}}\Delta S_{\text{Info}}=-T_{\text{E}}[\ln p_{\text{eq}}(r_i)-\ln p_{\text{eq}}(r_f)]=G_i-G_f+T_{\text{E}}\ln Z_i-T_{\text{E}}\ln Z_f$. Substituting this relation into the integral fluctuation theorem, we obtain the Jarzynski-like equality
\be
\ba
\label{eq9}
\langle \exp[-\beta_{\text{E}}(W+\Delta(PV))]\rangle=\exp(-\beta_{\text{E}}\Delta F_{\text{eff}}),
\ea
\ee
where the effective free energy is defined as $F_{\text{eff}}=-T_{\text{E}}\ln Z$. 
Although the mechanical work performed on the black hole remains $W=-\int PdV$, the Jarzynski relation involves the pressure-protocol work, $W+\Delta(PV)=\int VdP=\int \frac{\partial G}{\partial\lambda}d\lambda$, reflecting that the equilibrium landscape is governed by the Gibbs free energy rather than the internal energy. This choice is natural for black hole phase transitions in extended phase space, where the free-energy landscape governs the phase structure.

Applying Jensen's inequality to Eq.~(\ref{eq9}), followed by taking the logarithm and multiplying by $T_{\mathrm{E}}$, we obtain the second-law inequality
\be
\ba
\label{eq10}
-\langle W\rangle\leq -\Delta F_{\text{eff}}+\langle\Delta(PV)\rangle.
\ea
\ee
This inequality sets an upper bound on the average mechanical work extracted from the black hole in the extended phase space. The $PV$ term therefore acts as a boundary contribution that shifts the extractable work bound relative to that determined by the effective free-energy difference alone. In particular, a positive $\langle\Delta(PV)\rangle$ enlarges the upper bound on mechanical work extraction, whereas a negative value tightens it. Since the pressure is associated with the cosmological constant in the
extended phase space, the boundary contribution can be schematically
written as $\Delta(PV)=V_i\Delta P+P_f\Delta V$, where \(V_i\Delta P\) represents the contribution associated with the change in the cosmological constant, and hence in the AdS background geometry, evaluated at the initial thermodynamic volume, while \(P_f\Delta V\) characterizes the contribution arising from the change in the black-hole thermodynamic volume at the final pressure. Thus, the boundary contribution contains both the effect of the variation of the AdS background and that of the evolution of the black-hole configuration. Eq.~(\ref{eq10}) is actually the same as Eq.~(\ref{eq8}), we prove it as follows. The average information entropy is given by $\langle S_{\text{Info}}\rangle=-\int p_{eq}(r)\ln p_{eq}(r) dr=\beta_{\text{E}}\langle G\rangle+\ln Z$, where the equilibrium distribution $p_{eq}=\exp(-\beta_{\text{E}} G)/Z$ has been used. Substituting it into Eq.~(\ref{eq10}) to eliminate $T_{\mathrm{E}}\Delta\ln Z$, and then using $G=U+PV-T_{\text{E}}S_{\text{BH}}$, we obtain $\langle\Delta S_{\text{Info}}\rangle+\langle\Delta S_{\text{BH}}\rangle-\beta_{\text{E}}\langle\Delta U\rangle+\beta_{\text{E}}\langle W\rangle\geq 0$.
Since the energy conservation law $\langle\Delta U\rangle=\langle W\rangle+\langle Q\rangle$, the second law $\langle\Delta S_{\text{Info}}\rangle+\langle\Delta S_{\text{Bath}}\rangle+\langle\Delta S_{\text{BH}}\rangle\geq 0$ can be recovered.

\section{Numerical verification}

To numerically verify the Jarzynski-like equality, we design a protocol for varying the external pressure $P$ at a fixed ensemble temperature $T_{\text{E}}=0.03$. The total protocol duration is $\tau=8000$, with the initial pressure set to $P_i=0.406P_c$, at which the locally stable small and large black hole states have equal free energy. The system is initially prepared in equilibrium at $P_i$. During the first stage, $0 \le t \le \tau/5 = 1600$, the pressure is varied linearly to a final value $P_f$. In the subsequent stage, $1600 \le t \le \tau = 8000$, the pressure is held constant, and the system finally relaxes to an equilibrium state.

\begin{figure}[t]
\includegraphics[width=0.8\textwidth]{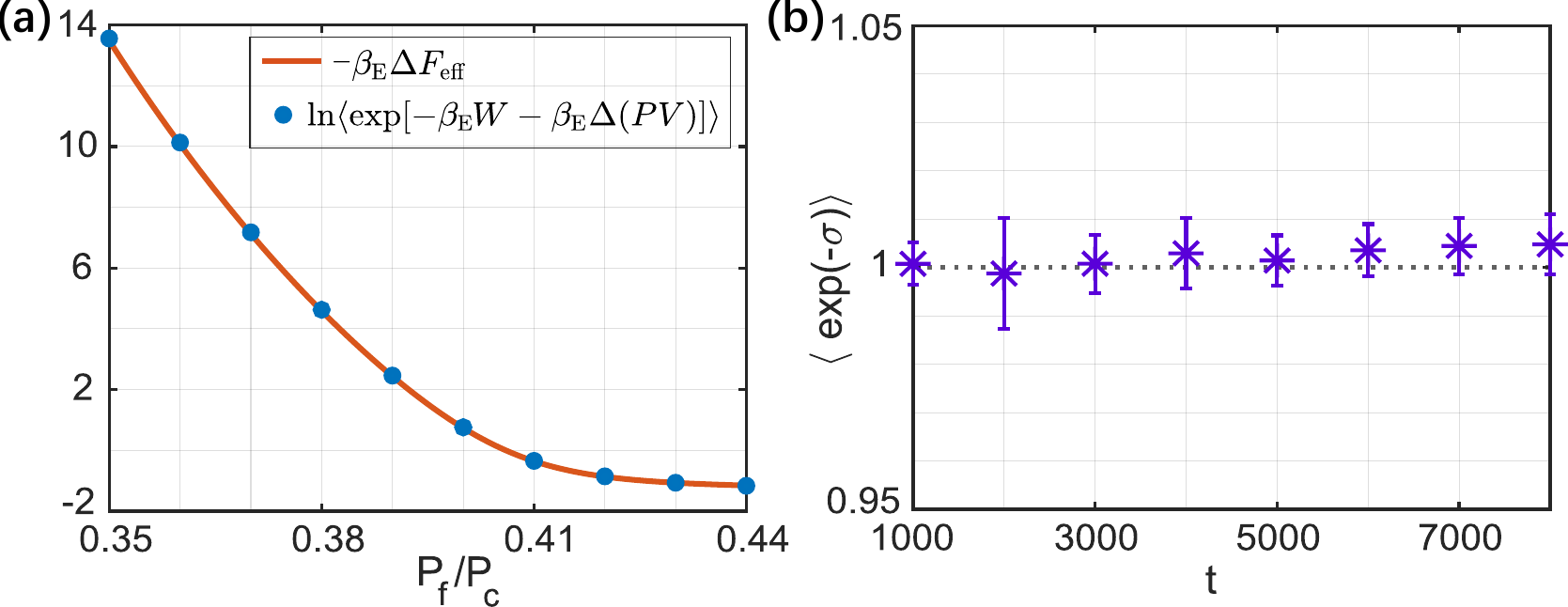}
\centering
\caption{(a) Numerical verification of the Jarzynski-like equality. The red line denotes the theoretical prediction evaluated from the initial and final equilibrium distributions, while the blue symbols represent numerical results obtained from 10000 stochastic trajectories. (b) Numerical verification of the integral fluctuation theorem.}
\label{Fig2}
\end{figure}

For each protocol characterized by a specific $P_f$, we compute the theoretical value $-\beta_{\mathrm{E}}\Delta F_{\mathrm{eff}}$ based on the initial and final equilibrium distributions, as shown by the red line in Fig.~\ref{Fig2} (a). Independently, we generate $10000$ stochastic trajectories with a time step of $0.001$, and evaluate $\exp[-\beta_{\mathrm{E}}(W+\Delta(PV))]$ along each trajectory. We then perform an ensemble average and take the logarithm of the result. The obtained values (blue points) show excellent agreement with the theoretical prediction (red line), thereby verifying the Jarzynski-like equality in Eq.~(\ref{eq9}), as illustrated in Fig.~\ref{Fig2}(a).

To verify the integral fluctuation theorem in Eq.~(\ref{eq7}), we do not require the initial and final states to be in equilibrium. Instead, we employ a fixed protocol with $P_f = 0.38 P_c$ and examine the validity of the theorem at different evolution times. For each case, we generate $16{,}000$ stochastic trajectories with a time step of $0.001$ and compute $\langle \exp(-\sigma)\rangle$. This procedure is repeated ten times to estimate the error bars shown in Fig.~\ref{Fig2}(b). Finally, we confirm the integral fluctuation theorem with deviations within $2\%$.

\section{A stochastic Stirling black hole heat engine}

In extended phase space thermodynamics, the use of AdS black holes as working substances to construct heat engine cycles in the pressure-volume plane was first proposed by C. V. Johnson and termed holographic heat engines~\cite{BU}. Previous studies have been restricted to classical thermodynamic cycles, where stochastic effects are absent and the efficiency is strictly bounded by the Carnot limit~\cite{BV,BW,BX,BY,BZ,CA,CB}. Here, we construct a stochastic Stirling black hole heat engine using the black hole dynamical phase transition and show that, at the level of single-trajectory realizations, the Carnot efficiency can be exceeded. However, upon ensemble averaging, the Carnot bound is recovered, even though the second law is modified in the form of Eq.~(\ref{eq8}).

\begin{figure}[t]
\includegraphics[width=0.8\textwidth]{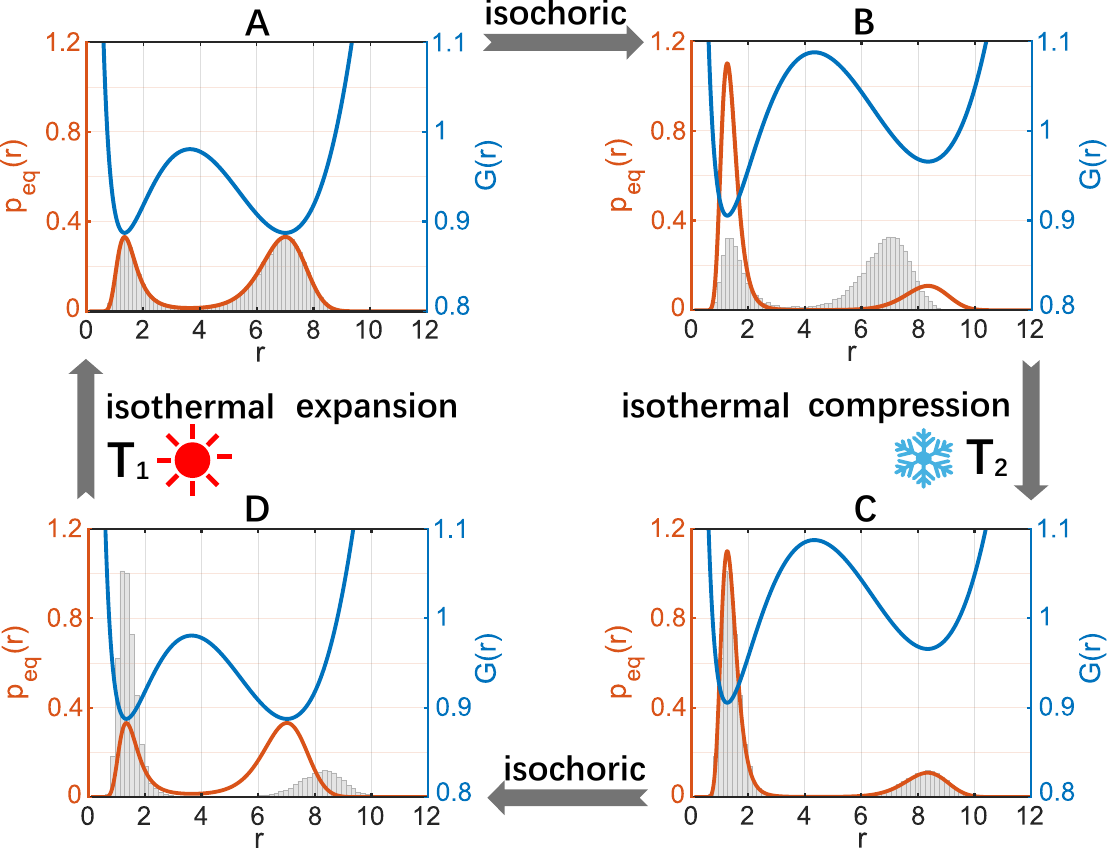}
\centering
\caption{A stochastic Stirling black hole heat engine. The blue curves represent the generalized free energy landscape, the red curves depict the corresponding equilibrium probability distributions, and the histograms indicate the actual instantaneous probability distributions at each labeled stage.}
\label{Fig3}
\end{figure}

Specifically, we construct a black hole heat engine model in the extended phase space, as illustrated schematically in Fig.~\ref{Fig3}. The thermodynamic cycle comprises four distinct processes. At the initial stage, labeled $A$, the system is prepared in the equilibrium configuration at ensemble temperature $T_1 = 0.03$ and pressure $P_1 = 0.406 P_c$, with the critical pressure $P_c=1/96\pi q^2$.

During the process $A \to B$, the ensemble temperature and pressure are instantaneously quenched to $T_2 = 0.026$ and $P_2 = 0.3 P_c$, respectively. Due to the sudden change, the system has no time to relax; consequently, the generalized free energy landscape undergoes an immediate transformation while the actual
instantaneous probability distribution remains frozen in its pre-quench configuration. As a result, the average volume is invariant and this stage is identified as an isochoric process.

In the subsequent process $B \to C$, the ensemble temperature and the pressure are held fixed at $T_2$ and $P_2$. The prevailing probability distribution at stage $B$ deviates from the equilibrium distribution dictated by the generalized free energy landscape, driving a stochastic relaxation toward equilibrium, during which the statistical weight of the small black hole configuration progressively increases. We consider the full-relaxation limit, in which the relaxation interval is sufficiently long for the distribution at stage \(C\) to reach the equilibrium configuration. For the specific parameter set used in Fig.~\ref{Fig3}, the average volume change is expressed as $\langle \Delta V_{\text{c}} \rangle = -419.84$, identifying this step as an isothermal compression in thermal contact with a cold bath. 

The process $C \to D$ involves a second instantaneous quench, restoring the ensemble temperature and pressure to their initial values $T_1$ and $P_1$. As before, the absence of a finite relaxation timescale renders this process isochoric.

During the process $D \to A$, the ensemble temperature and pressure remain fixed at $T_1$ and $P_1$. The system relaxes spontaneously toward a new equilibrium configuration, with the statistical weight of the large black hole state progressively increasing. We again consider the full-relaxation limit, so that the terminal distribution at stage $A$ returns to the equilibrium configuration. For the specific parameters illustrated, the average volumetric expansion is $\langle \Delta V_{\text{h}} \rangle  = 419.84$. Therefore, this process constitutes an isothermal expansion in contact with a hot bath. 

Upon completion of this final relaxation process, the system recovers the same equilibrium distribution from which the cycle originated, thereby closing the thermodynamic loop. For the full cycle, the second law in Eq.~(\ref{eq8}) constrains the average entropy production to be non-negative, now rewritten as: $\langle\sigma\rangle=\langle\Delta S_{\text{Bath}}\rangle=-{\langle Q_{\text{h}} \rangle}/{T_1}-{\langle Q_{\text{c}} \rangle}/{T_2}\geq 0$. Here, $\langle Q_{\text{h}} \rangle$ and $\langle Q_{\text{c}} \rangle$ denote the average heat absorbed from the hot and cold reservoirs, respectively. Consequently, the average efficiency of the heat engine is expressed as the ratio of the total extractable work $\langle W_{\text{ext}}\rangle$ to the heat absorbed from the hot bath: 
\be
\ba
\label{eq11}
\langle\eta\rangle=\frac{\langle W_{\text{ext}}\rangle}{\langle Q_{\text{h}} \rangle}=1+\frac{\langle Q_{\text{c}} \rangle}{\langle Q_{\text{h}} \rangle}\leq \eta_{\text{C}} = 1 - \frac{T_2}{T_1}, 
\ea
\ee
which is manifestly bounded from above by the Carnot efficiency $\eta_{\text{C}}$.

\begin{figure}[t]
\includegraphics[width=0.8\textwidth]{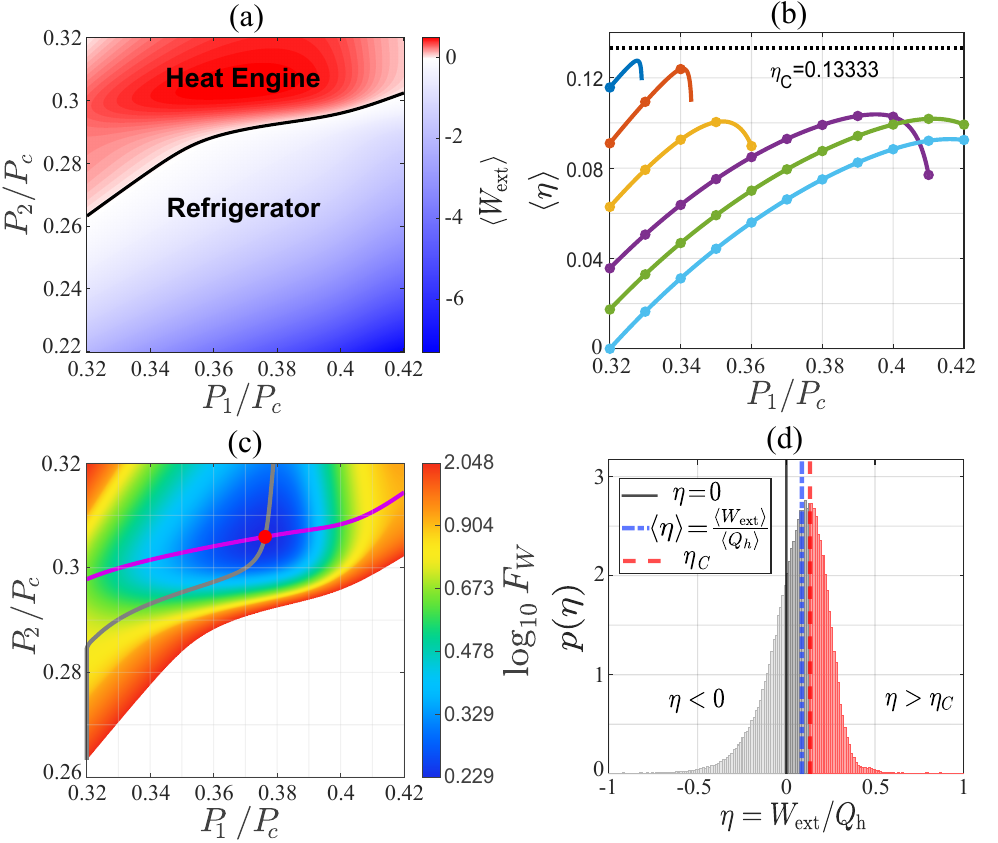}
\centering
\caption{(a) Operational duality of the thermodynamic cycle. (b) The ensemble-averaged efficiency \(\langle\eta\rangle\) remains bounded by the Carnot efficiency \(\eta_{\mathrm C}\). Solid curves from top to bottom correspond to \(P_2=[0.27,0.28,0.29,0.30,0.31,0.32]P_c\). 
(c) Fano factor \(F_W\) as a function of \(P_1\) and \(P_2\) within the heat-engine regime. The purple curve represents the minimum of \(F_W\) obtained by varying \(P_2\) at fixed \(P_1\), whereas the gray curve denotes the minimum obtained by varying \(P_1\) at fixed \(P_2\). The red dot marks the global minimum of \(F_W\), corresponding to the point of maximal operational stability. 
(d) Probability distribution of the stochastic efficiency in a single cycle evaluated at \(P_1=0.38P_c\) and \(P_2=0.31P_c\), corresponding to the regime with a suppressed work fluctuations.}
\label{Fig4}
\end{figure}

Keeping the reservoir temperatures fixed at $T_1 = 0.03$ and $T_2 = 0.026$, varying the external pressures $P_1$ and $P_2$ reveals a striking operational duality of the cycle, as shown in Fig.~\ref{Fig4}(a). For different choices of $(P_1, P_2)$, the same sequence of the four thermodynamic processes can function either as a heat engine or as a refrigerator. The latter, however, has received comparatively less attention in studies of black hole thermodynamics. Focusing on the heat-engine regime, we numerically evaluated the average efficiency for various $(P_1,P_2)$, and confirmed that it remains strictly below the Carnot bound $\eta_{\text{C}}$ in Fig.~\ref{Fig4}(b).

For a single-trajectory realization, the heat engine exhibits intrinsically stochastic behavior. To quantify the operational stability, we introduce the Fano factor of the extracted work, defined as \(F_W=({\langle W_{\mathrm{ext}}^2\rangle-\langle W_{\mathrm{ext}}\rangle^2})/\langle W_{\mathrm{ext}}\rangle\), which measures the work variance per unit mean extracted work. Thus, for a given average output, a smaller value of \(F_W\) indicates more suppressed work-output fluctuations and hence
corresponds to more stable engine performance. In Fig.~\ref{Fig4}(c), we plot \(F_W\) throughout the heat-engine regime as a function of \(P_1\) and \(P_2\). The optimal operating point, corresponding to the minimum fluctuation and hence the highest operational stability, is indicated by the red point.

In addition, the stochastic efficiency of the heat engine is defined as \(\eta=W_{\mathrm{ext}}/Q_{\text{h}}\). In sharp contrast to the ensemble-averaged efficiency, which is strictly bounded by the Carnot limit \(\eta_{\mathrm C}\), the trajectory-level efficiency is not fundamentally bounded. This originates from the fact that the stochastic entropy production may become negative along individual realizations, as required by the integral fluctuation theorem in Eq.~(\ref{eq7}). Consequently, transient super-Carnot efficiencies are not anomalous events, but rather an unavoidable manifestation of thermal fluctuations. As shown in Fig.~\ref{Fig4}(d), for the choice of \(P_1\) and \(P_2\) corresponding to a suppressed Fano factor, the probability distribution of \(\eta\) retains a finite weight in the regime \(\eta>\eta_{\mathrm C}\), highlighted in red. This demonstrates that, although the Carnot bound remains valid at the ensemble level, it can be transiently exceeded at the level of individual trajectories, thereby revealing a distinctive nonequilibrium advantage of stochastic black hole heat engines over their deterministic classical counterparts.

\section{Conclusions and discussions}

We have developed a stochastic thermodynamic framework for black hole phase transitions in extended phase space, thereby establishing a direct connection between conventional black hole thermodynamics and stochastic thermodynamics. At the trajectory level, the entropy production naturally decomposes into information, Bekenstein-Hawking, and bath contributions, leading to an integral fluctuation theorem and a generalized second law for fluctuating black holes. Our results demonstrate that, the conventional formulation of the second law, based only on the Bekenstein-Hawking entropy and bath contributions, is generically violated for the black hole phase transition unless the information entropy is explicitly included.

For protocols connecting equilibrium states, we further derived a Jarzynski-like equality in which the relevant work variable is $W+\Delta(PV)$ rather than the conventional $W$ alone. The $PV$ boundary term can enlarge the upper bound on the extractable mechanical work when $\langle\Delta(PV)\rangle>0$, relative to the bound set by the effective free-energy difference alone. In the reversible quasistatic limit, where this bound is saturated, a positive $\langle\Delta(PV)\rangle$ therefore contributes directly to the maximum extractable mechanical work in the extended phase space.

As a concrete application, we constructed a stochastic Stirling black hole engine and demonstrated that, although the Carnot bound is recovered upon ensemble averaging, individual trajectories can exceed it because negative entropy-production events remain allowed. This mechanism suggests a fluctuation-enabled route to enhanced performance of black hole heat engines via conditioning on physically admissible trajectories with negative stochastic entropy production. Our results provide a unified framework for fluctuations, irreversibility, and work relations in black hole thermodynamics, and open a route toward studying fluctuation-dominated processes in black-hole systems (such as primordial black holes in the early universe). The formalism can be extended to a broad class of black hole systems, including Schwarzschild-AdS, Gauss-Bonnet-AdS in modified gravity, and dyonic AdS black holes in the grand canonical ensemble with multiple order parameters, as detailed in the Appendix.

\section{Acknowledgements} 

C. H. Liu acknowledges support from the National Natural Science Foundation of China Grant Nos. 12547167 and 12234019. C. H. Liu also thanks the support by the Postdoctoral Fellowship Program and China Postdoctoral Science Foundation
under Grant Number GZC20261834.

\appendix
\section{Appendix: Extension to Diverse Black Hole Systems}

In the main text, we have established a stochastic thermodynamic framework for the charged AdS black hole. In this Appendix, we demonstrate that this formulation can be consistently extended to a broad class of gravitational systems, including the Schwarzschild-AdS (SAdS) black hole, Gauss-Bonnet-AdS (GBAdS) black holes in higher-curvature gravity, and dyonic AdS black holes in the grand canonical ensemble with multiple order parameters.

\subsection{SAdS Black Hole}

The metric of a $D$-dimensional SAdS black hole is given by
\be
\ba
ds^2 = - f(R)\,dt^2 + \frac{dR^2}{f(R)} + R^2 d\Omega_{D-2}^2,
\ea
\ee
where 
\be
\ba
f(R) = 1 - \frac{16\pi M}{(D-2)\Omega_{D-2} R^{D-3}} + \frac{R^2}{l^2}.
\ea
\ee

The thermodynamic quantities of the SAdS black hole are given by
\be
\ba
T_{\text{H}} =& \frac{1}{4\pi r}
\left[(D-3) + (D-1)\frac{r^2}{l^2}\right],
\qquad S_{\text{BH}}= \frac{\Omega_{D-2}}{4} r^{D-2},\\ 
M =& \frac{(D-2)\Omega_{D-2}}{16\pi} 
\, r^{D-3}
\left(1 + \frac{r^2}{l^2}\right),
\qquad
U=\frac{(D-2)\Omega_{D-2}}{16\pi} 
\, r^{D-3},\\
P=&\frac{(D-1)(D-2)}{16\pi l^2},\qquad 
V = \left(\frac{\partial M}{\partial P}\right)_S
= \frac{\Omega_{D-2}}{D-1} r^{D-1},
\ea
\ee
Thus, the generalized free energy is written as:
\be
\ba
G=M-T_{\text{E}}S_{\text{BH}}=\frac{(D-2)\Omega_{D-2}}{16\pi} 
\, r^{D-3}
\left(1 + \frac{r^2}{l^2}\right)-T_{\text{E}}\frac{\Omega_{D-2}}{4} r^{D-2},
\ea
\ee

The SAdS black hole exhibits a first-order phase transition between thermal AdS space and a stable large-black-hole phase. Dynamical descriptions based on the Fokker-Planck equation have been investigated previously~\cite {AT,AU}, while non-Markovian Langevin formulations have also been proposed~\cite{BD}. Here, we focus on a Markovian stochastic dynamics driven by Gaussian white noise $\eta(r,t)$. Treating the horizon radius $r$ as an effective order parameter, the Langevin equation is expressed as
\be
\ba
\label{eq2s}
\frac{dr}{dt}=-D\beta_E\,\partial_r G(r,\lambda)+\eta(r,t).
\ea
\ee
This dynamics admits an equivalent path-integral representation in terms of the Onsager-Machlup functional. The conditional probability of a forward trajectory starting from $r_i$ is given by
\be
\ba
p_{\text{F}}(r_{\text{F}}(t)|r_i)=N \exp\{-\int_{0}^{t_f}[\frac{1}{4D}{(\frac{dr}{dt}+\frac{D\beta_{\text{E}} \partial G(r,\lambda)}{\partial r})^2}-\frac{D\beta_{\text{E}} \partial^2 G(r,\lambda)}{2\partial r^2}]dt\},
\ea
\ee
where $\mathcal{N}$ is a normalization constant. Correspondingly, the backward conditional path probability in the backward process is given by:
\be
\ba
p_{\text{B}}(r_{\text{B}}(t)|r_f)=N \exp\{-\int_{0}^{t_f}[\frac{1}{4D}{(-\frac{dr}{dt}+\frac{D\beta_{\text{E}} \partial G(r,\lambda)}{\partial r})^2}-\frac{D\beta_{\text{E}} \partial^2 G(r,\lambda)}{2\partial r^2}]dt\}.
\ea
\ee

Taking the ratio of forward and backward path probabilities, we obtain
\be
\ba
\ln\frac{p_{\text{F}}(r_{\text{F}}(t)|r_i)}{p_{\text{B}}(r_{\text{B}}(t)|r_f)}=-\frac{1}{T_{\text{E}}}\int_{0}^{t_f}\frac{\partial G(r,\lambda)}{\partial r}\frac{dr}{dt} dt.
\ea
\ee
To evaluate the integrand, we compute
\be
\ba
\frac{\partial G}{\partial r}dr=&[\frac{(D-2)(D-3)}{16\pi}\Omega_{D-2}r^{D-4}+\frac{(D-2)(D-1)}{16\pi l^2}\Omega_{D-2}r^{D-2}]dr\\
&-T_{\text{E}}\frac{(D-2)}{4}\Omega_{D-2}r^{D-3}dr\\
=&\frac{(D-2)}{16\pi}[(D-3)+\frac{(D-1)}{l^2}r^2]\Omega_{D-2}r^{D-4}dr-T_{\text{E}}\frac{(D-2)}{4}\Omega_{D-2}r^{D-3}dr\\
=&T_{\text{H}}dS_{\text{BH}}-T_{\text{E}}dS_{\text{BH}}.
\ea
\ee
Therefore, 
\be
\ba
-\frac{1}{T_{\text{E}}}\int_0^{t_f} \frac{\partial G}{\partial r}dr&=-\int_0^{t_f} \frac{T_{\text{H}}}{T_{\text{E}}}dS_{\text{BH}}+\Delta S_{\text{BH}}\\
&=\Delta S_{\text{Bath}}+\Delta S_{\text{BH}}.
\ea
\ee
Here, the heat absorbed by the black hole is then determined from the energy conservation law as $\delta Q=dU-\delta W=\frac{(D-2)(D-3)}{16\pi}\Omega_{D-2}r^{D-4}dr+\frac{(D-1)(D-2)}{16\pi l^2}\Omega_{D-2}r^{D-2}dr=T_{\text{H}}dS_{\text{BH}}$, where $\delta W=-PdV$ denotes the work performed on the black hole. Consequently, the change in entropy of the heat bath at the ensemble temperature $T_{\text{E}}$ is given by $\Delta S_{\text{Bath}}=-\frac{1}{T_{\text{E}}}\int T_{\text{H}} dS_{\text{BH}}$.

The total entropy production $\sigma$ quantifies the time irreversibility of the trajectory, which is expressed as the
logarithm of the ratio of forward and backward trajectory
probabilities:
\be
\ba
\sigma&=\ln\frac{p_{\text{F}}(r_{\text{F}}(t))}{p_{\text{B}}(r_{\text{B}}(t))}\\
&=-\frac{1}{T_{\text{E}}}\int_{0}^{t_f}\frac{\partial G(r,\lambda)}{\partial r}\frac{dr}{dt} dt+\ln p(r_i)-\ln p(r_f)
\\
&=\Delta S_{\text{Bath}}+\Delta S_{\text{BH}}+\Delta S_{\text{info}}.
\ea
\ee
As a direct consequence, the integral fluctuation theorem holds,
\be
\ba
\langle\exp[-\sigma]\rangle=\sum_{r_{\text{F}(t)}}p_{\text{F}}(r_{\text{F}}(t))\frac{p_{\text{B}}(r_{\text{B}}(t))}{p_{\text{F}}(r_{\text{F}}(t))}=1,
\ea
\ee
where the average is taken over all forward trajectories. This immediately implies the non-negativity of the mean entropy production, $\langle\sigma\rangle=\langle\Delta S_{\text{Bath}}+\Delta S_{\text{BH}}+\Delta S_{\text{info}}\rangle\geq 0$.

\subsection{Modified Gravity Theory: GBAdS Black Hole}

For the $D$-dimensional Gauss-Bonnet AdS black hole, the metric takes the form
\be
\ba
ds^2 = - f(R)\, dt^2 + \frac{dR^2}{f(R)} + R^2 d\Omega_{D-2}^2,
\ea
\ee
where
\be
\ba
f(R) = 1 + \frac{R^2}{2\alpha}
\left[
1 - \sqrt{
1 + \frac{64\pi \alpha M}{(D-2)\Omega_{D-2} R^{D-1}}
- \frac{4\alpha}{l^2}
- \frac{4\alpha q^2}{R^{2D-4}}
}
\right].
\ea
\ee

The thermodynamic quantities of the GBAdS black hole are given by
\be
\ba
P=&\frac{(D-1)(D-2)}{16\pi l^2},\qquad 
S_{\text{BH}}=\frac{1}{4}\Omega_{D-2}r^{D-2}\left(1+\frac{2\alpha(D-2)}{D-4}\frac{1}{r^2}\right),\\ 
V=&\frac{\Omega_{D-2}}{D-1} r^{D-1}, \qquad
U = \frac{(D-2)\Omega_{D-2}}{16\pi}
\left[
r^{D-3} + \alpha r^{D-5}
+ \frac{q^2}{ r^{D-3}}
\right],\\
T_{\text{H}}=&\frac{(D-3)r^2+(D-1)r^4/l^2+(D-5)\alpha-(D-3){q^2}/{r^{2D-8}}}{4\pi r(r^2+2\alpha)},\\
G=&M-T_{\text{E}}S\\
=&\frac{(D-2)\Omega_{D-2}}{16\pi}
\left(
r^{D-3} + \alpha r^{D-5}
+ \frac{16\pi P}{(D-1)(D-2)} r^{D-1}
+ \frac{q^2}{ r^{D-3}}
\right)\\
&- \frac{1}{4}T_{\text{E}}\Omega_{D-2}r^{D-2}
\left(1+\frac{2\alpha(D-2)}{D-4}\frac{1}{r^2}\right),
\ea
\ee

The order parameter is chosen to be the horizon radius, and the system is therefore still described within the canonical ensemble. Owing to the presence of a single order parameter, the Langevin equation takes the same form as that for the SAdS black hole. Using the Onsager-Machlup path integral formalism, we obtain
\be
\ba
\ln\frac{p_{\text{F}}(r_{\text{F}}(t)|r_i)}{p_{\text{B}}(r_{\text{B}}(t)|r_f)}=-\frac{1}{T_{\text{E}}}\int_{0}^{t_f}\frac{\partial G(r,\lambda)}{\partial r}\frac{dr}{dt} dt.
\ea
\ee

Evaluating $-\frac{\partial G(r)}{\partial r}dr$, we find
\be
\ba
-\frac{\partial G(r)}{\partial r}dr=&-\frac{1}{16\pi}(D-2)\Omega_{D-2}\left[(D-3)r^{D-4}+\frac{16\pi}{(D-2)}Pr^{D-2}+(D-5)\alpha r^{D-6}\right.\\
&\left.+(3-D)q^2r^{2-D}\right]dr+\frac{1}{4}T_E\Omega_{D-2}\left((D-2)r^{D-3}+2\alpha(D-2)r^{D-5}\right)dr
\ea
\ee
Note that
\be
\ba
-T_{\text{H}}dS_{\text{BH}}=&-\frac{(D-3)r^2+(D-1)r^4/l^2+(D-5)\alpha-(D-3)q^2/r^{2D-8)}}{4\pi r(r^2+2\alpha)}\\
&\times\frac{1}{4}\Omega_{D-2}(D-2)r^{D-5}(r^{2}+2\alpha)dr\\
=&-\frac{1}{16\pi}(D-2)\Omega_{D-2}\left[(D-3)r^{D-4}+\frac{16\pi }{(D-2)}Pr^{D-2}+(D-5)\alpha r^{D-6}\right.\\
&\left.- (D-3)q^2r^{2-D}\right]dr,
\ea
\ee
\be
\ba
T_{\text{E}}dS_{\text{BH}}=\frac{1}{4}\Omega_{D-2}T_E(D-2)r^{D-5}(r^{2}+2\alpha)\,dr,
\ea
\ee
and thus
\be
\ba
-\frac{\partial G(r)}{\partial r}dr=-T_{\text{H}}dS_{\text{BH}}+T_{\text{E}}dS_{\text{BH}}.
\ea
\ee
The heat absorbed by the black hole is then determined from the energy conservation law as
\be
\ba
\delta Q=&dU-\delta W=\frac{(D-2)\Omega_{D-2}}{16\pi}[(D-3)r^{D-4}+\alpha(D-5)r^{D-6}+q^2(3-D)r^{2-D}]dr\\
&+\frac{(D-1)(D-2)}{16\pi l^2}\Omega_{D-2}r^{D-2}dr\\
=&\frac{(D-3)r^2+(D-1)r^4/l^2+(D-5)\alpha-(D-3){q^2}/{r^{2D-8}}}{4\pi r(r^2+2\alpha)}\\
&\times \frac{(D-2)}{4}\Omega_{D-2}r^{D-5}(r^2+2\alpha)dr\\
=&T_{\text{H}}dS_{\text{BH}}.
\ea
\ee
Consequently, the entropy change of the heat bath is given by $\Delta S_{\text{Bath}}=-\frac{1}{T_{\text{E}}}\int T_{\text{H}} dS_{\text{BH}}$.

The total entropy production $\sigma$ is then expressed as
\be
\ba
\sigma=\ln\frac{p_{\text{F}}(r_{\text{F}}(t))}{p_{\text{B}}(r_{\text{B}}(t))}=\Delta S_{\text{Bath}}+\Delta S_{\text{BH}}+\Delta S_{\text{info}}.
\ea
\ee
The integral fluctuation theorem holds,
\be
\ba
\langle\exp[-\sigma]\rangle=\sum_{r_{\text{F}(t)}}p_{\text{F}}(r_{\text{F}}(t))\frac{p_{\text{B}}(r_{\text{B}}(t))}{p_{\text{F}}(r_{\text{F}}(t))}=1,
\ea
\ee
which implies that the average entropy production is nonnegative, $\langle\sigma\rangle=\langle\Delta S_{\text{Bath}}+\Delta S_{\text{BH}}+\Delta S_{\text{info}}\rangle\geq 0$.

\subsection{Grand Canonical Ensemble: Dyonic AdS Black Hole}

For the four-dimensional dyonic AdS black hole, the metric is given by:
\be
\ba
ds^2=-f(R)dt^2+\frac{1}{f(R)}dR^2+R^2d\theta^2+R^2\sin^2\theta d\phi^2,
\ea
\ee
with
\be
\ba
\label{eq:4}
f(R)=1+\frac{R^2}{l^2}-\frac{2M}{R}+\frac{q_e^2+q_m^2}{R^2}.
\ea
\ee
where $q_e$ and $q_m$ are the electric and magnetic charges, respectively.

The thermodynamic quantities of the dyonic AdS black hole are written as
\be
\ba
M=&\frac{r}{2}+\frac{r^3}{2l^2}+\frac{q_e^2+q_m^2}{2r},\qquad
U=\frac{r}{2}+\frac{q_e^2+q_m^2}{2r},\\
P=&\frac{3}{8\pi l^2},\qquad
T_{\text{H}}=\frac{1}{4\pi r}(1+\frac{3r^2}{l^2}-\frac{q_e^2+q_m^2}{r^2}),
\ea
\ee
In the grand canonical ensemble with fixed electric potential $\Phi_e$ (and fixed magnetic charge $q_m$), the Gibbs free energy is given by
\be
\ba
G=U+PV-T_{\text{E}}S_{\text{BH}}-\Phi_{e}q_e
=\left(\frac{r}{2}+\frac{q_e^2+q_m^2}{2r}\right)+\frac{r^3}{2l^2}-T_{\text{E}}(\pi r^2)-\Phi_e q_e.
\ea
\ee

The stochastic dynamics of the phase transition are described by an overdamped Langevin equation for the order parameters $\vec{x}=(r,q_e)$. To avoid notational confusion, we introduce a constant, isotropic diffusion coefficient $D$ and the inverse bath temperature $\beta_{\text{E}} = 1/T_{\text{E}}$. The Langevin equation then reads
\be
\ba
\label{eq:40}
\frac{d\vec{x}}{dt}=-{D}\beta_{\text{E}}\nabla G(\vec{x})+\vec{\eta}(\vec{x},t),
\ea
\ee
or, in components,
\be
\ba
\label{eq:41}
&\frac{dr}{dt}=-\frac{{D}\beta_{\text{E}}\partial G(r,q_e)}{\partial r}+\eta_1(r,q_e,t),\\
&\frac{dq_e}{dt}=-\frac{{D}\beta_{\text{E}}\partial G(r,q_e)}{\partial q_e}+\eta_2(r,q_e,t).
\ea
\ee
The Gaussian white noises $\eta_i(r,q_e,t)$ satisfy
\be
\ba
\label{eq:42}
<\eta_i(r,q_e,t)>=0,\quad <\eta_i(r,q_e,t)\eta_j(r,q_e,0)>=2D\delta(t)\delta_{ij},\quad i,j=1,2.
\ea
\ee

The conditional probability of a forward trajectory $\vec{x}_{\text{F}}(t)$ is given by the Onsager-Machlup functional
\be
\ba
p_{\text{F}}[\vec{x}_{\text{F}}(t)|\vec{x}_i]=N\exp\{-\int[\frac{(\frac{dr}{dt}-f_r)^2}{4D}+\frac{(\frac{dq_e}{dt}-f_{q_e})^2}{4D}+\frac{\partial_rf_r+\partial_{q_e}f_{q_e}}{2}]dt\},
\ea
\ee
while the conditional probability of the backward trajectory $\vec{x}_{\text{B}}(t)$ is
\be
\ba
p_{\text{B}}[\vec{x}_{\text{B}}(t)|\vec{x}_f]=N\exp\{-\int[\frac{(-\frac{dr}{dt}-f_r)^2}{4D}+\frac{(-\frac{dq_e}{dt}-f_q)^2}{4D}+\frac{\partial_rf_r+\partial_{q_e}f_{q_e}}{2}]dt\},
\ea
\ee
where $(f_r,f_{q_e})=(-\beta_{\text{E}} D\partial_r G, -\beta_{\text{E}} D\partial_{q_e} G)$. From these expressions, one obtains the ratio between forward and backward path probabilities as
\be
\ba
\ln\frac{p_{\text{F}}[\vec{x}_{\text{F}}(t)|\vec{x}_i]}{p_B[\vec{x}_{\text{B}}(t)|\vec{x}_f]}=-\frac{1}{T_{\text{E}}}
\left[ \int_{\vec{x}_i}^{\vec{x}_f}
\nabla G(\vec{x})\cdot d\mathbf{x}\right].
\ea
\ee

Then, we evaluate $-\nabla G(r,q_e) d{\bf x}$, and its explicit form is
\be
\ba
\label{eq:33}
-\nabla G(r,q_e) d{\bf x}=&-\frac{\partial G(r,q_e)}{\partial r}dr-\frac{\partial G(r,q_e)}{\partial q_e}dq_e\\
=&-\left[\frac{1}{2}-\frac{q_e^2+q_m^2}{2r^2}+\frac{3r^2}{2l^2}-2\pi T_{\text{E}} r\right]dr+\left(-\frac{q_e}{r}+\Phi_e\right) dq_e\\
=&-2\pi r T_{\text{H}} dr+2\pi T_{\text{E}}rdr+\left(-\frac{q_e}{r}+\Phi_e\right) dq_e\\
=&\left(-T_{\text{H}}dS_{\text{BH}}+\Phi_e dq_e-\Phi_{\text{H}}dq_e\right)+T_{\text{E}}dS_{\text{BH}},
\ea
\ee
where we used $S_{\text{BH}} = \pi r^2$, hence $dS_{\text{BH}} = 2\pi r dr$, and the horizon electric potential $\Phi_{\text{H}} = q_e/r$.

Using $dU=\delta Q -PdV+\Phi_{\text{H}} dq_e$, we have
\be
\ba
\left(\frac{1}{2}-\frac{q_e^2+q_m^2}{2r^2}\right)dr+\frac{q_e}{r}dq_e=\delta Q- P4\pi r^2 dr+\frac{q_e}{r}dq_e,
\ea
\ee
which gives
\be
\ba
\delta Q=\left(\frac{1}{2}-\frac{q_e^2+q_m^2}{2r^2}+4P\pi r^2\right) dr=2\pi T_{\text{H}} r dr=T_{\text{H}}dS_{\text{BH}}.
\ea
\ee

From the perspective of the bath, we consider the energy exchange between the system and the total environment (the heat bath plus the particle bath). The energy change of the system is
\be
\ba
dU_{\text{BH}}=T_{\text{H}}dS_{\text{BH}}-PdV+\Phi_{\text{H}}dq_e,
\ea
\ee
and the energy change of the bath is
\be
\ba
dU_{\text{Bath}}=T_EdS_{\text{Bath}}+PdV-\Phi_edq_e.
\ea
\ee
Here, the $PdV$ term represents the mechanical work received by the environment due to the volume change of the black hole. Since the black hole and the environment are treated as an energy-conserving composite system, this mechanical work contribution enters the energy balances of the two subsystems with opposite signs. Energy conservation, $-dU_{\text{BH}}=dU_{\text{Bath}}$, gives
\be
\ba
-T_{\text{H}}dS_{\text{BH}}+PdV-\Phi_{\text{H}}dq_e=T_{\text{E}}dS_{\text{Bath}}+PdV-\Phi_edq_e,
\ea
\ee
hence
\be
\ba
dS_{\text{Bath}}=-\frac{T_{\text{H}}}{T_{\text{E}}}dS_{\text{BH}}+\frac{\Phi_e-\Phi_{\text{H}}}{T_{\text{E}}}dq_e.
\ea
\ee
It is important to note that, in the presence of particle exchange, the heat exchanged between the black hole and the thermal environment does not generally satisfy the simple relation $T_H dS_{\rm BH}=-T_E dS_{\rm Bath}$. The deviation arises because part of the energy exchange is associated with particle transfer under a chemical-potential difference, as reflected by the chemical work term.

With these results, the integrand of the path-probability ratio can be rewritten as
\be
\ba
-\frac{\nabla G(r,q_e)}{T_{\text{E}}}d{\bf x}=dS_{\text{Bath}}+dS_{\text{BH}}.
\ea
\ee
This directly leads to an integral fluctuation theorem and a non-negative entropy production.

The total entropy production $\sigma$ is defined as
\be
\ba
\sigma=\ln\frac{p_{\text{F}}(\vec{r}_{\text{F}}(t))}{p_{\text{B}}(\vec{r}_{\text{B}}(t))}=\Delta S_{\text{Bath}}+\Delta S_{\text{BH}}+\Delta S_{\text{info}},
\ea
\ee
where $\Delta S_{\text{info}}=-\ln p(\vec{x}_f)+\ln p(\vec{x}_i)$. The integral fluctuation theorem then follows:
\be
\ba
\langle\exp[-\sigma]\rangle=\sum_{r_{\text{F}(t)}}p_{\text{F}}(r_{\text{F}}(t))\frac{p_{\text{B}}(r_{\text{B}}(t))}{p_{\text{F}}(r_{\text{F}}(t))}=1,
\ea
\ee
which in turn guarantees that the average entropy production is non-negative: $\langle\sigma\rangle=\langle\Delta S_{\text{Bath}}+\Delta S_{\text{BH}}+\Delta S_{\text{info}}\rangle\geq 0$.


\begin{thebibliography}{99}

\bibitem{AA}
J. M. Bardeen, B. Carter, and S. W. Hawking, \emph{The four laws of black hole mechanics}, Commun. Math. Phys. {\bf 31}, 161 (1973).
\bibitem{AB}
J. D. Bekenstein, \emph{Black holes and entropy}, Phys. Rev. D {\bf { 7}}, 2333 (1973).
\bibitem{AC}
S. W. Hawking, \emph{Particle creation by black holes}, Commun. Math. Phys. {\bf 43}, 199 (1975).
\bibitem{AD}
S. W. Hawking, \emph{Black Holes and Thermodynamics}, Phys. Rev. D {\bf 13}, 191 (1976).
\bibitem{AE}
D. Kastor, S. Ray, and J. Traschen, \emph{Enthalpy and the Mechanics of AdS Black Holes}, Class. Quant. Grav. {\bf 26}, 195011 (2009).
\bibitem{AF}
B. P. Dolan, \emph{The cosmological constant and the black hole equation of state}, Class. Quant. Grav. {\bf 28}, 125020 (2011).
\bibitem{AG}
B. P. Dolan, \emph{Pressure and volume in the first law of black hole thermodynamics}, Class. Quant. Grav. {\bf 28}, 235017 (2011).
\bibitem{AH}
D. Kubiznak and R. B. Mann, \emph{P-V criticality of charged AdS black holes}, J. High Energy Phys. {\bf 1207}, 033 (2012).
\bibitem{AI}
D. Kubiznak, R. B. Mann, and M. Teo, \emph{Black hole chemistry: thermodynamics with Lambda}, Class. Quant. Grav. {\bf 34}, 063001 (2017).
\bibitem{AJ}
R.B.Mann, \emph{Black hole chemistry: The first 15 years}, Int. J. Mod. Phys. D {\bf 34}, 2542001 (2025).
\bibitem{AK}
N. Altamirano, D. Kubiznak, and R. B. Mann, \emph{Reentrant phase transition in rotating anti-de Sitter black holes}, Phys. Rev. D {\bf 88}, 101502 (2013).
\bibitem{AL}
S. W. Wei and Y. X. Liu, \emph{Triple points and phase diagrams in the extended phase space of charged Gauss-Bonnet black holes in AdS space}, Phys. Rev. D {\bf 90}, 044057 (2014).
\bibitem{AM}
M. Tavakoli, J. Wu, and R. B. Mann, \emph{Multi-critical Points in Black Hole Phase Transitions}, J. High Energy Phys. {\bf 12}, 117 (2022).
\bibitem{AN}
R. A. Hennigar, R. B. Mann, and E. Tjoa, \emph{Superfluid Black Holes}, Phys. Rev. Lett. {\bf 118}, 021301 (2017).
\bibitem{AO}
 K. Sekimoto,
 \emph{Stochastic Energetics} (Springer, Berlin Heidelberg, 2010).
 \bibitem{AP}
 U. Seifert, \emph{Stochastic thermodynamics, fluctuation theorems and molecular machines}, Rep. Prog. Phys. {\bf 75}, 126001 (2012).
\bibitem{AQ}
U. Seifert, \emph{Entropy production along a stochastic trajectory and an integral fluctuation theorem}, Phys. Rev. Lett. {\bf 95}, 040602 (2005).
\bibitem{AR}
C. Jarzynski, \emph{Nonequilibrium Equality for Free Energy Differences}, Phys. Rev. Lett. {\bf 78}, 2690 (1997).
\bibitem{AS}
D. J. Searles and D. J. Evans, \emph{The fluctuation theorem and Green-Kubo relations}, J. Chem. Phys. {\bf 112}, 9727 (2000).
%
\bibitem{GGB}
J. Garriga and A. Vilenkin, \emph{ Cycles in the Multiverse}, Phys. Rev. D {\bf 85}, 103509 (2012).
\bibitem{GGC}
D. Schwartz-Perlov and A. Vilenkin, \emph{Probabilities in the Bousso-Polchinski multiverse}, JCAP {\bf 06}, 010 (2006).
\bibitem{GGD}
J. Garriga, D. Schwartz-Perlov, A. Vilenkin and S. Winitzki, \emph{Probabilities in the inflationary
multiverse}, JCAP {\bf 01}, 017 (2006).
\bibitem{GGE}
H. Wang, X. Y. Li, and J. Wang,
\emph{Quantifying the potential and flux landscapes for nonequilibrium multiverse, a new scenario for time arrow}, J. High Energy Phys. {\bf 02}, 105 (2021).
\bibitem{AT}
R. Li and J. Wang, \emph{Thermodynamics and kinetics of Hawking-Page phase transition}, Phys. Rev. D {\bf 102} (2020) 024085.
\bibitem{AU}
R. Li, K. Zhang, and J. Wang, \emph{Thermal dynamic phase transition of Reissner-Nordstr$\ddot{o}$m Anti-de Sitter black holes on the free energy landscape}, J. High Energy Phys. {\bf 10} (2020) 090.
%
\bibitem{AV}
R. G. Cai, \emph{Oscillatory behaviors near a black hole triple
point}, Sci. China Phys. Mech. Astron. {\bf 64}, 290432 (2021).
\bibitem{AW}
S. W. Wei, Y. Q. Wang, Y. X. Liu, and R. B. Mann, \emph{Observing dynamic oscillatory behavior of triple points among black hole thermodynamic phase transitions}, Sci. China Phys. Mech. Astron. {\bf 64}, 270411 (2021).
%
\bibitem{AX}
Z. M. Xu, \emph{Fokker-Planck equation for black holes in thermal potential}, Phys. Rev. D {\bf 104}, 104022 (2021).
\bibitem{AY}
S. W. Wei, Y.X. Liu, and Y. Q. Wang, \emph{Dynamic properties of thermodynamic phase transition for five-dimensional neutral Gauss-Bonnnet AdS black hole on free energy landscape}, Nucl. Phys. B {\bf 976}, 115692 (2022).
\bibitem{AZ}
S. J. Yang, R. Zhou, S. W. Wei, and Y. X. Liu, \emph{Kinetics of a phase transition for a Kerr-AdS black hole on the free-energy landscape}, Phys. Rev. D {\bf 105}, 084030 (2022).
\bibitem{BA}
J. Y. Yang and R. B. Mann, \emph{Dynamic behaviours of black hole phase transitions near quadruple points}, J. High Energy Phys. {\bf 08},  028 (2023).
%
\bibitem{BBBBA}
T. Q. Yue, J. Wang, \emph{Transit time of black holes on generalized free energy landscape}, J. High Energy Phys. {\bf 06},  080 (2025).
%
\bibitem{BB}
M. A. S. Afshar, S. N. Gashti, M. R. Alipour, J. Sadeghi, \emph{Overcoming Barriers: Kramers' escape rate analysis of metastable dynamics
in first-order multi-phase transitions}, J. High Energy Phys. {\bf 11}, 163 (2025).
%
\bibitem{BC}
M. S. Ali, H. El Moumni, J. Khalloufi and K. Masmar, \emph{Revisiting the Born-Infeld-AdS black hole phase structure through Landau theory and free energy landscape approaches}, Sci Rep. {\bf 15}, 15425 (2025).
\bibitem{ZZZZN}
Z. M. Xu, B. Wu, and W. L. Yang, \emph{Rate of the phase transition for a charged anti-de Sitter black hole}, Sci. China Phys. Mech. Astron. {\bf 66}, 240411 (2023).
\bibitem{ZZZZO}
Y. S. Wang, Z. M. Xu, and B. Wu, \emph{Thermodynamic phase transition rate for the third-order Lovelock black hole in diverse dimensions}, Phys. Lett. B {\bf 853}, 138690 (2024).
\bibitem{ZZZZP}
C. Ma, P. P. Zhang, B. Wu, and Z. M. Xu, \emph{The Kramers escape rate of phase transitions for the 6-dimensional Gauss-Bonnet AdS black hole with triple phases}, Phys. Lett. B {\bf 861}, 139282 (2025).
\bibitem{ZZZZQ}
C. Ma, B. Wu, and Z. M. Xu, \emph{Mean first passage time and the Kramers escape rate of phase transitions for the Bardeen-AdS-class black hole}, Chin. Phys. C {\bf 50}, 085103 (2026).
\bibitem{BD}
R. Li and J. Wang, \emph{Kinetics of Hawking-Page phase transition with the non-Markovian effects}, J. High Energy Phys. {\bf 05}, 128 (2022).
\bibitem{BE}
R. Li, C. H. Liu, K. Zhang, and J. Wang, \emph{Topology of the landscape and dominant kinetic path for the thermodynamic phase transition of the charged Gauss-Bonnet-AdS black holes}, Phys. Rev. D {\bf 108}, 044004 (2023).
\bibitem{BF}
C. H. Liu and J. Wang, \emph{Path integral and instantons for the dynamical process and phase transition rate of Reissner-Nordstr$\ddot{o}$m-AdS black holes}, Phys. Rev. D {\bf 105}, 104024 (2022).
\bibitem{BG}
C. Liu, R. Li, K. Zhang, and J. Wang, \emph{Generalized free energy and dynamical state transition of the dyonic AdS black hole in the grand canonical ensemble}, J. High Energy Phys. {\bf 11}, 068 (2023).
\bibitem{BH}
R. Li and J. Wang, \emph{Generalized free energy landscape of black hole phase transition}, Phys. Rev. D {\bf 106}, 106015 (2022).
\bibitem{BI}
S. P. Wu, Y. X. Liu, and S. W. Wei, \emph{Generalized free energy landscapes from Iyer-Wald formalism}, Phys. Dark Univ. {\bf 51}, 102210 (2026).
%
\bibitem{BJ}
S. W. Wei, Y. X. Liu, and R. B. Mann, \emph{Black Hole Solutions as Topological Thermodynamic Defects}, Phys. Rev. Lett. {\bf 129}, 191101 (2022).
%
\bibitem{BK}
S. W. Wei, Y. X. Liu, and R.B. Mann, \emph{Universal topologi
cal classifications of black hole thermodynamics}, Phys. Rev.
D {\bf 110}, L081501 (2024).
%
\bibitem{BL}
D. Wu and S. Q. Wu, \emph{Topological classes of thermodynamics of rotating AdS black holes}, Phys. Rev. D {\bf 107}, 084002 (2023).
%
\bibitem{BM}
D. Wu, S. Y. Gu, X. D. Zhu, Q. Q. Jiang, and S. Z. Yang,
\emph{Topological classes of thermodynamics of the static multi
charge AdS black holes in gauged supergravities: Novel
temperature-dependent thermodynamic topological phase
transition}, J. High Energy Phys. {\bf 06}, 213 (2024).
%
\bibitem{BN}
C.X. Fang, J. Jiang, and M. Zhang, \emph{Revisiting thermodynamic
topologies of black holes}, J. High Energy Phys. {\bf 01}, 102 (2023).
%
\bibitem{BO}
C. H. Liu and J. Wang, \emph{Topological natures of the Gauss-Bonnet black hole in AdS space}, Phys. Rev. D {\bf 107}, 064023 (2023).
%
\bibitem{BP}
 J. Sadeghi, S. Noori Gashti, M.R. Alipour, and M.A.S. Afshar, \emph{Bardeen black hole thermodynamics from topological perspective}, Annals Phys. {\bf 455}, 169391 (2023).
 %
 \bibitem{BQ}
D. Chen, Y. He and J. Tao, \emph{Topological classes of higher-dimensional black holes in massive gravity}, Eur. Phys. J. C {\bf 83}, 872 (2023).
 %
 \bibitem{BR}
M. S. Ali, H. El Moumni, J. Khalloufi and K. Masmar, \emph{Topology of Born-Infeld-AdS black hole phase transitions: Bulk and CFT sides}, Annals Phys. 465, 169679 (2024).
%
\bibitem{BS}
H. S. Wio, \emph{Path Integrals for Stochastic Processes: An Introduction} (World Scientific Press, Singapore, 2013).
\bibitem{BT}
L. Onsager and S. Machlup, \emph{Fluctuations and irreversible processes}, Phys. Rev. {\bf 91} (1953) 1505.
%
\bibitem{SSIEQ}
In the Langevin equation, the drift force is given by the gradient of a generalized free energy and is therefore conservative, implying that the steady state corresponds to an equilibrium distribution.
%
\bibitem{BU}
C. V. Johnson, \emph{Holographic Heat Engines}, Class. Quant. Grav. {\bf 31}, 205002 (2014).
%
\bibitem{BV}
J. L. Zhang, Y. J. Li, and H. W. Yu, \emph{Thermodynamics of
charged accelerating AdS black holes and holographic heat
engines}, J. High Energy Phys. {\bf 02}, 144 (2019).
%
\bibitem{BW}
J. X. Mo, F. Liang, G. Q. Li, \emph{Heat engine in the three-dimensional spacetime}, J. High Energy Phys. {\bf 03},  
010 (2017).
%
\bibitem{BX}
C. Bhamidipati, P. K. Yerra, \emph{Heat engines for dilatonic Born-Infeld black holes}, Eur. Phys. J. C {\bf 77}, 534 (2017).
%
\bibitem{BY}
S. W. Wei and Y. X. Liu, \emph{Charged AdS black hole
heat engines}, Nucl. Phys. B {\bf 946}, 114700 (2019).
%
\bibitem{BZ}
C. V. Johnson, \emph{Exact model of the power-to-efficiency trade-off while approaching the Carnot limit}, Phys. Rev. D 98, 026008 (2018).
%
\bibitem{CA}
M. C. DiMarco, S. L. Jess, R. A. Hennigar, R. B. Mann, \emph{Universality for black hole heat engines near critical points}, Phys. Rev. D {\bf 107}, 044001 (2023).
%
\bibitem{CB}
R. A. Hennigar, F. McCarthy, A. Ballon and R.B. Mann, \emph{Holographic heat engines: general
considerations and rotating black holes}, Class. Quant. Grav. {\bf 34}, 175005 (2017).

\end{thebibliography}
\end{document}